\documentclass[11pt]{article}
\pdfoutput=1
\usepackage[centertags]{amsmath}
\usepackage[square, comma, sort&compress,numbers]{natbib}
\usepackage{array,multirow}
\numberwithin{equation}{section}
\usepackage{amssymb,amsfonts}
\usepackage{graphicx}
\usepackage{color}
\usepackage{mathtools}

\usepackage{epsfig}
\usepackage{bbold}

\usepackage{wrapfig}
\usepackage{float}
\usepackage{soul}
\usepackage{tkz-euclide}

\usepackage{tikz,pgf}
\usetikzlibrary{shapes}
\usetikzlibrary{calc}
\usetikzlibrary{decorations.pathmorphing}
\usetikzlibrary{decorations.pathreplacing,shapes.misc}
\usetikzlibrary{positioning}
\usetikzlibrary{arrows}
\usetikzlibrary{decorations.markings}
\usetikzlibrary{shadings}
\usetkzobj{all}

\usetikzlibrary{intersections}

\def\be{\begin{equation}}
\def\ee{\end{equation}}
\def\ba{\begin{array}}
\def\ea{\end{array}}

\def\dps{\displaystyle}
\newcommand{\half}{\frac{1}{2}}

\def\1{\tilde{1}}
\def\2{\tilde{2}}
\def\3{\tilde{3}}

\newdimen\tableauside\tableauside=1.0ex
\newdimen\tableaurule\tableaurule=0.4pt
\newdimen\tableaustep
\def\phantomhrule#1{\hbox{\vbox to0pt{\hrule height\tableaurule
width#1\vss}}}
\def\phantomvrule#1{\vbox{\hbox to0pt{\vrule width\tableaurule
height#1\hss}}}
\def\sqr{\vbox{%
  \phantomhrule\tableaustep

\hbox{\phantomvrule\tableaustep\kern\tableaustep\phantomvrule\tableaustep}%
  \hbox{\vbox{\phantomhrule\tableauside}\kern-\tableaurule}}}
\def\squares#1{\hbox{\count0=#1\noindent\loop\sqr
  \advance\count0 by-1 \ifnum\count0>0\repeat}}
\def\tableau#1{\vcenter{\offinterlineskip
  \tableaustep=\tableauside\advance\tableaustep by-\tableaurule
  \kern\normallineskip\hbox
    {\kern\normallineskip\vbox
      {\gettableau#1 0 }%
     \kern\normallineskip\kern\tableaurule}%
  \kern\normallineskip\kern\tableaurule}}
\def\gettableau#1 {\ifnum#1=0\let\next=\null\else
  \squares{#1}\let\next=\gettableau\fi\next}

\newcommand{\bref}[1]{\textbf{\ref{#1}}}

\def\cF{\mathcal{F}}

\def\cL{\mathcal{L}}

\def\cO{\mathcal{O}}

\numberwithin{equation}{section} \makeatletter
\@addtoreset{equation}{section}

\def\be{\begin{equation}}
\def\ee{\end{equation}}
\def\ba{\begin{array}}
\def\ea{\end{array}}

\def\dps{\displaystyle}

\def\ba{\begin{array}}
\def\ea{\end{array}}

\def\dps{\displaystyle}

\def\cbeta{\beta}

\def\rads{R}

\def\lengthB{\cL_{_{\hspace{-0.5mm}AdS_{_3}[3]}}}

\def\banados{Ba\~{n}ados }
\def\mobius{M\"obius\;}
\def\LH{L$^k$H$^{n-k}$ }
\def\cft{CFT$_2$ }

\usepackage{jheppub}
\makeatletter
\def\@fpheader{\vspace{-.1cm}}
\makeatother

\title{Holographic variables for CFT$_2$ conformal blocks with heavy  operators}

\author[a,b]{Konstantin\ Alkalaev}

\author[a]{Mikhail\ Pavlov}

\affiliation[a]{I.E. Tamm Department of Theoretical Physics, \\P.N. Lebedev Physical
Institute,\\ Leninsky ave. 53, 119991 Moscow, Russia}
\affiliation[b]{Department of General and Applied Physics, \\
Moscow Institute of Physics and Technology, \\
Institutskiy per. 7, Dolgoprudnyi, \\141700 Moscow region, Russia}
\emailAdd{alkalaev@lpi.ru}
\emailAdd{pavlov@lpi.ru}

\abstract{ We consider large-$c$ $n$-point Virasoro blocks  with $n-k$ background heavy operators and $k$ perturbative heavy operators. Conformal dimensions of heavy operators scale linearly with large $c$, while splitting into background/perturbative operators assumes an additional perturbative expansion. Such conformal blocks can be  calculated within the monodromy method that basically reduces to solving auxiliary Fuchsian second-order equation and finding monodromy of solutions. We show that there exist particular variables that we call holographic, use of which drastically simplifies the whole analysis. In consequence, we formulate the uniformization property of the large-$c$ blocks which states that in the holographic variables their form depends only on the number of perturbative heavy operators. On the other hand, the  holographic variables encode the metric in the bulk space so that the conformal blocks with the same number of perturbative operators are calculated by the same geodesic trees but on  different geometries created by the background operators.}

\makeatletter
\def\@fpheader{\vspace{-.1cm}}
\makeatother

\begin{document}

\maketitle
\flushbottom

\section{Introduction}


Virasoro conformal block functions  $\cF(x|\Delta, \tilde \Delta,c)$ \cite{Belavin:1984vu} are not known in closed form for general values of conformal dimensions $\Delta, \tilde \Delta$ and the central charge $c$. On the other hand, the AdS/CFT correspondence motivates the study of the regime when the central charge tends to infinity, $c\to \infty$.  If external and intermediate conformal dimensions $\Delta, \tilde \Delta$ are {\it heavy}, i.e. they scale linearly with large $c$, then the original conformal block takes a simpler, exponential form  \cite{Zamolodchikov1986}. However, such large-$c$ conformal blocks with heavy operators are still quite complicated functions. In the bulk, the Brown-Henneaux relation \cite{Brown:1986nw} says that the large-$c$ blocks can be reproduced  from the three-dimensional quantum gravity path integral evaluated in the semiclassical approximation.

Further simplification can be achieved by considering the so-called heavy-light expansion \cite{Fitzpatrick:2014vua}, when a number of original heavy primary operators forms a background for other heavy primary operators, i.e. $\Delta_p/\Delta_b \ll 1$, where $\Delta_b$ and $\Delta_p$ are dimensions of the background and perturbative operators, respectively. The resulting perturbative conformal blocks are much simpler as compared to the original large-$c$ blocks. From the holographic perspective, the perturbative blocks are calculated by lengths of geodesic trees stretched in the bulk space created by the background heavy operators \cite{Asplund:2014coa,Fitzpatrick:2014vua,Hijano:2015rla,Fitzpatrick:2015zha,Alkalaev:2015wia,Hijano:2015qja,Banerjee:2016qca,Alkalaev:2016rjl}.

Let \LH denote $n$-point perturbative conformal block with $n-k$ background heavy operators and $k$ perturbative heavy operators. The most studied cases include the 4-point LLHH blocks \cite{Asplund:2014coa,Fitzpatrick:2014vua,Hijano:2015rla,Fitzpatrick:2015zha,Hijano:2015qja,Fitzpatrick:2016mtp}, the 5-point LLLHH blocks \cite{Alkalaev:2015wia,Alkalaev:2015lca,Alkalaev:2015fbw,Belavin:2017atm}, the $n$-point L$^{n-2}$HH blocks \cite{Banerjee:2016qca,Alkalaev:2016rjl,Alkalaev:2018nik}, and 4-point LHHH block \cite{Alkalaev:2019zhs}.

In this paper, we continue the study of \LH perturbative conformal blocks by revealing previously  hidden structure that underlies the heavy-light expansion. By that we mean that the perturbative blocks  allow for  very special parameterization that we call {\it holographic variables}. Since $n$ coordinates  of $n$-point perturbative blocks are naturally split into two parts, one can transform coordinates of the perturbative operators by means of a particular mapping function, while keeping coordinates of the background operators intact. The mapping function can be explicitly defined by using solutions to the auxiliary Fuchsian equation. It is parameterized by coordinates of the background operators. Such a transformation allows to reorganize the original coordinate dependence of perturbative blocks so that now they depend on the holographic variables only. In fact, the holographic variables realize the general observation of  \cite{Fitzpatrick:2015zha} that, owing to fact that the stress tensor is not primary, the dependence on the background operators can be absorbed by performing a particular conformal transformation.

A remarkable consequence of using the holographic variables is the {\it  uniformization} of  perturbative conformal  blocks already discussed in \cite{Fitzpatrick:2015zha,Fitzpatrick:2016mtp,Anous:2019yku} in the case of two background operators. For $n$-point  blocks, it may be formulated as follows: the  perturbative blocks of L$^k$H$^{n-k}$ and L$^k$H$^{m-k}$ types being represented in terms of the holographic variables have the same form at $m\neq n$. From the holographic perspective, the uniformization   is quite natural. Indeed, the background operators define the bulk space while the perturbative operators are realized via dual geodesic trees. The shape of geodesic trees is defined by perturbative operators only and not by the background operators.


The outline of this paper is as follows. In Section \bref{sec:mon} we discuss the monodromy method and formulate the heavy-light expansion which finally defines $n$-point \LH perturbative blocks. In Section \bref{sec:HoloV} we introduce the holographic variables and formulate the uniformization  property of the perturbative blocks. Section \bref{sec:E} contains examples of LLHH and LLHHH blocks which demonstrate the use of the holographic variables. In Section \bref{sec:bulk} the holographic variables are explicitly  related to building the dual three-dimensional geometry created by the background operators. Here, using the holographic variables we identify a dual geodesic tree which length calculates the perturbative LLHHH block. Section \bref{sec:sum} summarizes our  findings.


\section{Classical conformal blocks and heavy-light expansion}
\label{sec:mon}

We consider holomorphic Virasoro $n$-point conformal block  $\cF(x|\Delta, \tilde \Delta,c)$ in a given OPE channel \cite{Belavin:1984vu}. Here, $x = \{x_1,.., x_n\}$ denotes coordinates  of primary operators with holomorphic conformal dimensions $\Delta$, intermediate holomorphic conformal dimensions are denoted by  $\tilde \Delta$, and $c$ is the central charge. Let all external and intermediate conformal dimensions be heavy, i.e. grow linearly with the central charge, $\Delta = \cO(c)$ and $\tilde \Delta = \cO(c)$. In the large-$c$ regime the conformal block behaves exponentially \cite{Zamolodchikov1986,Besken:2019jyw}
\be
\label{classical}
\cF(x| \Delta, \tilde \Delta,c) \,\Big |_{c\to\infty} \;\rightarrow\;\; \exp\big[\,\frac{c}{6}f(x| \epsilon, \tilde \epsilon)\,\big]\;, \quad
\epsilon_i = \frac{6\Delta_i}{c}\;,\quad
\tilde \epsilon = \frac{6\tilde\Delta}{c}\;,
\ee
where $f(x| \epsilon, \tilde \epsilon)$ is the classical conformal block which depends on the central charge only through the classical dimensions $\epsilon, \tilde \epsilon$.

A convenient way to calculate large-$c$ conformal blocks is the monodromy method.\footnote{For review and recent studies of the monodromy method  see e.g. \cite{Harlow:2011ny,Hartman:2013mia,Fitzpatrick:2014vua,Hijano:2015rla,Alkalaev:2015lca,Anous:2016kss,Alkalaev:2016rjl,Anous:2017tza,Kusuki:2018nms}.} To this end, one considers an auxiliary $(n + 1)$-point conformal block with an additional degenerate operator of light conformal dimension $\cO(c^0)$. Due to the fusion rules  the auxiliary block in the large-$c$ regime factorizes as
\be
\psi(y|x) \exp\big[\,\frac{c}{6}f(x| \epsilon, \tilde \epsilon)\,\big]\;,
\ee
where $f(x|\epsilon,\tilde\epsilon)$ is $n$-point classical block \eqref{classical} and $\psi(y|x)$ stands for the large-$c$ contribution of the degenerate operator.

Imposing the BPZ condition one obtains the Fuchsian type equation \cite{Belavin:1984vu}
\be
\label{BPZ}
\left[\frac{d^2}{dy^2}  + T(y|x)\right]\psi(y|x) = 0\;,
\qquad
T(y|x) = \sum_{m=1}^n \frac{\epsilon_m}{(y-x_m)^2} + \frac{c_m}{y-x_m}\;,
\quad
c_m = \frac{\partial f(x| \epsilon, \tilde \epsilon)}{\partial x_m} \;,
\ee
with $n$ singular points given by positions of the original primary operators.
Here, the function $T(y|x)$ is the stress tensor, the gradients $c_m$ are the accessory parameters which can be found  by studying the monodromy properties of the Fuchsian equation \eqref{BPZ} (see below). Note that there are three constraints
\be
\label{linear}
\sum_{m=1}^n c_m = 0\;,
\qquad
\sum_{m=1}^n (c_m x_m+ \epsilon_m) = 0\;,
\qquad
\sum_{m=1}^n (c_m x^2_m+ 2\epsilon_m x_m) = 0\;,
\ee
ensuring that the algebraic part of \eqref{BPZ} has no singularity at $y\to\infty$. Knowing all the accessory parameters one can integrate the gradient equations to obtain the classical block.

\paragraph{Heavy-light expansion.} Finding classical blocks can be drastically simplified by employing  the so-called heavy-light expansion \cite{Fitzpatrick:2014vua}. Suppose now that $n-k$ heavy operators with classical dimensions $\epsilon_j$ are much heavier than other $k$ heavy operators,
\be
\label{perl}
\epsilon_i \ll \epsilon_j\;, \qquad i = 1,..,k \;,\qquad j = k+1,..,n \;.
\ee
Then, the positions of all operators can be split into two subsets: perturbative sector and background sector  $x = \{ z \,, \,{\bf z}\} \equiv  \{z_1,.., z_k, {\bf z}_{k+1},.., {\bf z}_{n}\}$.

Now, we implement the heavy-light expansion
\be
\label{decos}
\begin{gathered}
\psi(y| z , {\bf z}) = \psi^{(0)}(y|{\bf z}) + \psi^{(1)}(y| z , {\bf z})+...\,,
\qquad
T(y|z,  {\bf z}) = T^{(0)}(y|{\bf z}) + T^{(1)}(y|z , {\bf z})+...\,,
\\
f(z, {\bf z}|\epsilon, \tilde{\epsilon}) = f^{(0)}({\bf z}|\epsilon, \tilde{\epsilon}) + f^{(1)}(z, {\bf z}|\epsilon, \tilde{\epsilon})  + ...\,,
\qquad
c_{m}(z , {\bf z}|\epsilon, \tilde{\epsilon}) = c_{m}^{(0)}({\bf z}|\epsilon,\tilde{\epsilon}) + c_{m}^{(1)}(z , {\bf z}|\epsilon, \tilde{\epsilon})  + ...\,,
 \end{gathered}
\ee
where $m = 1,...,n$. By construction, the zeroth-order accessory parameters of the perturbative operators are zero, $c_{i}^{(0)} = 0$, $i=1,...,k$.

The constraints  \eqref{linear} can be  expanded similarly as
\be
\label{linear0}
\sum_{j=k+1}^n c^{(0)}_j = 0\;,
\qquad
\sum_{j=k+1}^n (c^{(0)}_j {\bf z}_j+ \epsilon_j) = 0\;,
\qquad
\sum_{j=k+1}^n (c^{(0)}_j {\bf z}^2_j+ 2\epsilon_j {\bf z}_j) = 0\;,
\ee
\be
\label{linear1}
\sum_{m=1}^n c^{(1)}_m = 0\;,
\qquad
\sum_{m=1}^n c^{(1)}_m z_m+ \sum_{i=1}^k\epsilon_i = 0\;,
\qquad
\sum_{m=1}^n c^{(1)}_m z^2_m+ \sum_{i=1}^k 2\epsilon_i z_i = 0\;.
\ee

\vspace{1mm}

\paragraph{Zeroth-order solutions.} In the zeroth-order, the Fuchsian equation \eqref{BPZ} takes the form
\be
\label{fuchs0}
\left[\frac{d^2}{dy^2}  + T^{(0)}(y|{\bf z})\right]\psi(y|{\bf z}) = 0\;,
\quad
\text{where}
\quad T^{(0)}(y|{\bf z}) = \sum_{j=k+1}^n \frac{\epsilon_j}{(y-{\bf z}_j)^2} + \frac{c^{(0)}_j}{y-{\bf z}_j}\;,
\ee
and its solutions are given by two independent branches
\be
\label{sol0}
\psi^{(0)}_{\pm} = \psi^{(0)}_{\pm}(y|{\bf z},\epsilon, c^{(0)})\;.
\ee
Here,  $c_j^{(0)}$, $j = k+1,...,n$, are independent parameters that can be found by solving constraints \eqref{linear0} and the gradient equations
\be
c_j^{(0)} = \frac{\partial f^{(0)}({\bf z}|\epsilon, \tilde{\epsilon}) }{\partial {\bf z}_j}\;,
\qquad j = k+1,...,n\;.
\ee
Since the background conformal block is assumed to be known, then $c_j^{(0)}$ can be found explicitly and substituted back into \eqref{sol0} to obtain $\psi^{(0)}_{\pm} =
\psi^{(0)}_{\pm}(y|{\bf z},\epsilon)$.\footnote{\label{foot} In this form, the solutions $\psi^{(0)}_{\pm}$ are explicitly known for two \cite{Fitzpatrick:2014vua} and three \cite{Alkalaev:2019zhs} background operators, see also Section \bref{sec:E}.} Usually, the linear constraints are solved in the very beginning to isolate $n-3$ independent accessory parameters. However, equally one can keep $n$ accessory parameters independent and solve the three constraints \eqref{linear} at later stages. In that case, solutions to the Fuchsian equation in the zeroth order are explicitly parameterized by the background accessory parameters \eqref{sol0}.

Two comments are in order. First, the solution \eqref{sol0} near singular points ${\bf z}_j$ behaves as
\be
\label{sing}
\psi^{(0)}_{\pm} \sim  (y-{\bf z}_j)^{\frac{1\pm \alpha_j}{2}}\;,
\qquad
\alpha_j = \sqrt{1-4 \epsilon_j}\;,
\qquad
j=k+1,...\,,n\;,
\ee
that follows from that the leading asymptotics are defined by the most singular terms in \eqref{fuchs0}. The exponents are restricted as\footnote{\label{foot1} For two background operators with dimension $\epsilon_{n-1} =\epsilon_n \equiv \epsilon_h$ the range \eqref{alpha}  corresponds to conical singularities in $AdS_3$, whereas for $\epsilon_h \geq  \frac14$ we have a (threshold)  BTZ black hole \cite{Fitzpatrick:2014vua, Asplund:2014coa}. In this case, the solutions \eqref{sing} are analytically continued to purely imaginary $\alpha$. For three and  more background operators, one would  expect that the range $\epsilon_j \geq \frac14$ would correspond to multi BTZ-like  solutions though their explicit form and  general properties are yet unknown (see, however, discussions in  \cite{Brill:1995yc,Coussaert:1994if,Barbot:2005qk,Mansson:2000sj}). Also, for many background operators, it is possible to consider a combination of conical singularities and BTZ black holes.}.
\be
\label{alpha}
0 < \epsilon_j < \frac14\;,
\qquad
0<\alpha_j <1\;.
\ee
Second, the zeroth-order solutions are hard to find for any number of heavy background insertions except for two and three background operators in which case the Fuchsian equation can be solved explicitly (see the footnote \bref{foot}). More than three background operators require the knowledge of higher-point classical conformal  blocks $f^{(0)}({\bf z}|\epsilon, \tilde{\epsilon})$, which can be calculated only as power series in coordinates ${\bf z}$.

\paragraph{First-order solutions.} In the first-order  the equation \eqref{BPZ} is reduced to
\be
\label{T1}
\ba{c}
\dps \left[\frac{d^2}{dy^2} + T^{(0)}(y|{\bf z})\right]\psi^{(1)}(y|z, {\bf z}) = - T^{(1)}(y|z,  {\bf z}) \psi^{(0)}(y|{\bf z})\;, \\
\\
\dps T^{(1)}(y|z,  {\bf z})  = \sum_{i=1}^{k} \left(\frac{\epsilon_i}{(y-z_i)^2} + \frac{c^{(1)}_i}{y-z_i} \right) + \sum_{j=k+1}^n \frac{c^{(1)}_j}{y-{\bf z}_j}\;.
\ea
\ee
The solution is given in terms of the zeroth-order solutions \eqref{sol0} as
\be
\label{sol1}
\ba{l}
\dps \psi^{(1)}_{\pm}(y|z, {\bf z}) = \frac{1}{W({\bf z})}\left(\psi^{(0)}_{+}(y|{\bf z})
\int dy \; \psi^{(0)}_{-}(y|{\bf z}) T^{(1)}(y|z, {\bf z})\psi^{(0)}_{\pm}(y|{\bf z})  \right.
\\
\\
\dps
\hspace{50mm}\left.-\psi^{(0)}_{-}(y|{\bf z})\int dy \; \psi^{(0)}_{+}(y|{\bf z}) T^{(1)}(y|z, {\bf z})\psi^{(0)}_{\pm}(y|{\bf z})\right)\;,
\ea
\ee
where the Wronskian is
\be
\label{wronskian0}
W({\bf z}) \equiv - \psi^{(0)}_{+}(y|{\bf z}) \frac{d \psi^{(0)}_{-}(y|{\bf z})}{dy} + \psi^{(0)}_{-}(y|{\bf z})  \frac{d \psi^{(0)}_{+}(y|{\bf z})}{dy}\;.
\ee
The Wronskian is independent of $y$ that is the general property of Fuchsian equations.

\paragraph{Monodromy analysis.} The monodromy method consists of comparing the monodromy of solutions to the Fuchsian equation against that of the original correlation function. This yields a system of algebraic equations on the accessory parameters. In principle, the system can be solved and then the problem of finding the classical block can be reduced to solving the gradient equations \eqref{BPZ}.

To this end, let us consider contours $\Gamma_p$ encircling points $\{z_1,..., z_{k}, {\bf z}_{k+1},..., {\bf z}_{p+1}\}$, where $p = 1,...,n-3$.  The monodromy matrices along  $\Gamma_p$ are defined as
\be
\psi_a(\Gamma_p \circ y|z,{\bf z}) = M_{ab}(\Gamma_p|z,{\bf z}) \psi_b(y|z,{\bf z})\;,
\qquad
a,b = \pm\;,
\ee
and, within the heavy-light expansion, the monodromy matrices can be decomposed as
\be
\label{dec}
M_{ab}(\Gamma_p|z,{\bf z}) = M_{ab}^{(0)} (\Gamma_p|{\bf z}) + M_{ab}^{(1)}(\Gamma_p|z,{\bf z}) +...\;,
\ee
where $M_{ab}^{(0)} (\Gamma_p|{\bf z})$ is defined by the zeroth-order solution  \eqref{sol0},   $M_{ab}^{(1)}(\Gamma_p|z,{\bf z})$ is defined by the first-order solution \eqref{sol1}. Due to the form of \eqref{sol1} the first-order correction  factorizes as
\be
\label{mf1}
M_{ab}^{(1)}(\Gamma_p|z,{\bf z}) = - M_{ac}^{(0)} (\Gamma_p|{\bf z})  I_{cb}(\Gamma_p|z,{\bf z})\;, \qquad  I_{cb}  = \begin{pmatrix}
  I^{(p)}_{++}\;\;& I^{(p)}_{+-}\\
 I^{(p)}_{-+}\;\;&  I^{(p)}_{--}
\end{pmatrix} \,,
\ee
where
\be
\label{int}
\ba{c}
\dps
I^{(p)}_{+\pm}(z, {\bf z})=  \frac{1}{W({\bf z})}\int_{\Gamma_p} dy \; \psi^{(0)}_{+}(y|{\bf z}) T^{(1)}(y|z, {\bf z}) \psi^{(0)}_{\mp}(y|{\bf z})\;,
\\
\\
\dps
I^{(p)}_{-\mp}(z, {\bf z}) = -\frac{1}{W({\bf z})}\int_{\Gamma_p} dy \; \psi^{(0)}_{\pm}(y|{\bf z}) T^{(1)}(y|z, {\bf z}) \psi^{(0)}_{-}(y|{\bf z})\;.
\ea
\ee
The above integrals are straightforward to calculate since the stress tensor $T^{(1)}$ \eqref{T1} has a simple pole structure, and
\be
\label{assim}
\psi^{(0)}_{-}\psi^{(0)}_{-} \sim (y-{\bf z}_j)^{1-\alpha_j}\;,
\qquad
\psi^{(0)}_{+}\psi^{(0)}_{+} \sim (y-{\bf z}_j)^{1+\alpha_j}\;,
\qquad
\psi^{(0)}_{-}\psi^{(0)}_{+} \sim (y-{\bf z}_j)\;,
\ee
with the exponents satisfying \eqref{alpha}.

On the other hand, traversing the light degenerate operator $V_{{(2,1)}}(y)$ in the original $(n+1)$-point correlation function along contours $\Gamma_p$ we find the respective monodromy matrices
\be
\label{pres}
\widetilde{M}_p = - \begin{pmatrix} e^{i \pi  \gamma_p}& 0\\
  0& e^{-i \pi   \gamma_p}
\end{pmatrix}\,,\qquad
\gamma_p = \sqrt{1- 4\tilde{\epsilon}_p}\;,
\qquad p = 1,..., n - 3\;.
\ee
Equating the  eigenvalues of these matrices with those of \eqref{mf1} yields a system of $n-3$ algebraic equations on perturbative accessory parameters. Recalling that there are three additional constraints  \eqref{linear1} we conclude that in total there are $n$ equations on $n$ accessory parameters.


\section{Holographic variables}
\label{sec:HoloV}


Let us consider {\it the  holographic function} and its derivative defined as
\be
\label{phv}
w(y|{\bf z}) = \frac{\psi_{+}^{(0)}(y|{\bf z})}{\psi_{-}^{(0)}(y|{\bf z})}\;, \qquad w'(y|{\bf z}) = \frac{W({\bf z})}{\left(\psi_{-}^{(0)}(y| {\bf z})\right)^2}\;,
\ee
where $\psi_{\pm}^{(0)}(y|{\bf z})$ are solutions to the zeroth-order Fuchsian equation   \eqref{sol0}  and the prime denotes a derivative with respect to $y$. The second relation follows from the first one  by virtue of   \eqref{wronskian0}. Note that function $w(y|{\bf z})$ is determined up to the \mobius transformation since in \eqref{phv} we can equally take linear combinations of solutions. Recalling \eqref{sing} we find that the functions \eqref{phv} behave near the singular points ${\bf z}_j$ as
\be
\label{sing_w}
w(y|{\bf z}) \sim (y-{\bf z}_j)^{\alpha_j}\;,
\qquad
w'(y|{\bf z}) \sim (y-{\bf z}_j)^{\alpha_j-1}\;,
\ee
where the exponents are restricted by \eqref{alpha}.

Now, we consider a partial conformal map such that coordinates of the perturbative operators are replaced by values of the holographic function $w(y|{\bf z})$, i.e.
\be
\label{partial}
\{z_1,..., z_k, {\bf z}_{k+1},..., {\bf z}_{n}\}\; \rightarrow \;\; \{w(z_1|{\bf z}),...,w(z_k|{\bf z}),  {\bf z}_{k+1},..., {\bf z}_{n}\}\;.
\ee
We leave the coordinates of the background operators intact, otherwise $w({\bf z}_j|{\bf z}) = 0$, $j=k+1,...,n$ due to  the singular behaviour \eqref{sing_w}. Evaluating  functions \eqref{phv} at $y=z_i$ we denote
\be
\label{hv}
w_{i} \equiv w(z_i|{\bf z})\;,
\qquad\;\; i = 1,... ,k\;.
\ee
The values $w_i$ can be called {\it holographic coordniates} because of the special role they play in the dual bulk geometry (see Section \bref{sec:GT}). Equivalently, the holographic function \eqref{phv} defines the map of $k$-dimensional complex spaces $\mathbb{C}^k \to \mathbb{C}^k$, which is parameterized by ${\bf z}$. This map is invertible. Indeed, the Jacobi matrix is diagonal $J_{ij} = w'_i \,\delta_{ij}$, where $w'_i=w'(z_i|{\bf z})$ are derivatives \eqref{phv} evaluated at $y=z_i$. Since $w'(y|{\bf z})$  can have zeros/poles only  at points $y={\bf z}_j$ \eqref{sing_w}, then the Jacobi matrix is non-degenerate.

\paragraph{Monodromy integrals.} Using the holographic variables the monodromy integrals along contours $\Gamma_p$ \eqref{int} can be represented as 
\be
\ba{c}
\label{HI}
\dps I^{(p)}_{+-}(z, {\bf z}) = \int_{\Gamma_p} dy\, \frac{ w^2(y|{\bf z}) }{w'(y|{\bf z})}\,T^{(1)}(y|z, {\bf z})\;, \qquad I^{(p)}_{-+}(z, {\bf z}) = -\int_{\Gamma_p} dy \, \frac{1}{w'(y|{\bf z})}\,T^{(1)}(y|z, {\bf z})\;, \\
\\
\dps I^{(p)}_{++}(z, {\bf z}) = -I^{(p)}_{--}(z, {\bf z})= \int_{\Gamma_p} dy\, \frac{ w(y|{\bf z}) }{w'(y, {\bf z})}\,T^{(1)}(y|z, {\bf z})\;,
\ea
\ee
and explicitly calculated by means of  the residue theorem,
\be
\ba{l}
\label{GS}
\dps \frac{I^{(p)}_{++}}{2 \pi i} = \sum^{min\{p+1,k\}}_{i=1} \left(X_i w_i+\epsilon_i\right)\,,
\\
\\
\dps \frac{I^{(p)}_{-+}}{2 \pi i} = -\sum^{min\{p+1,k\}}_{i=1} X_i \,,
\\
\\
\dps \frac{I^{(p)}_{+-}}{2 \pi i} = \sum^{min\{p+1,k\}}_{i=1} \left(X_i w^2_i+ 2 \epsilon_i w_i\right)\,,
\ea
\ee
where instead of original first-order accessory parameters we introduced
\be
\label{def_par}
X_i = \frac{1}{w'_i}\left(c^{(1)}_i- \epsilon_i \frac{w''_i}{w'_i}\right)\;,
\quad
i = 1,...,k\;,
\qquad \;\;
Y_j = c^{(1)}_{j}\;,
\quad
j=k+1,...,n\;.
\ee

A few comments are in order. Firstly, it is crucial that the upper limit value $min\{p+1,k\}$ leads to that all integrals over contours $\Gamma_p, \; p \geq k - 1$  are equal to $I^{(k)}_{\pm\pm}$. This is why the integrals are independent of the accessory parameters $Y_j$ \eqref{def_par}. Secondly, the monodromy integrals explicitly depend on the holographic variables only, while dependence on $z_i$ and ${\bf z}_j$ is implicit. Thirdly, the integrals are simple linear functions of new parameters $X_i$ and  remarkably mimic the linear constraints \eqref{linear}.

\paragraph{Zeroth-order monodromy.} Now, comparing eigenvalues of the monodromy matrices \eqref{mf1} and \eqref{pres}  in  the zeroth order yields the conditions
\be
\label{fusion}
\epsilon_{j+1} = \tilde{\epsilon}_{j}\;, \qquad j = k+1,...,n-3 \;.
\ee
It means that the heavy-light expansion is possible only if all pairs of adjacent  external and intermediate dimensions in the background part of the original $n$-point classical block $f(z, {\bf z}|\epsilon, \tilde{\epsilon})$ are equated so that the background block $f^{(0)}({\bf z}|\epsilon, \tilde{\epsilon})$ is not general. It means that some of expansion coefficients in coordinates of $f(z, {\bf z}|\epsilon, \tilde{\epsilon})$ contain poles in perturbative dimensions $\epsilon_i$ with prefactors $(\epsilon_{j+1} - \tilde{\epsilon}_j)$.

\paragraph{First-order monodromy.} In the first order, the monodromy equations take the form
\be
\label{mon1}
I^{(p)}_{++}I^{(p)}_{++} + I^{(p)}_{+-} I^{(p)}_{-+} = - 4\pi^2 \tilde{\epsilon}^2_p\;,
\qquad p = 1,...,k-1\;,
\ee
\be
\label{mon2}
I^{(p)}_{++} = I^{(k-1)}_{++}=0\;,
\qquad p = k,...,n-3\;.
\ee 
We see that there are linearly dependent equations in \eqref{mon2} so that in total  there are $k$ independent equations in \eqref{mon1}, \eqref{mon2} for $k$ variables $X_i$, $i=1,...,k$. It follows that the monodromy equations allow  to find  the accessory parameters of the perturbative operators only, i.e. $X_i = X_i(w|\epsilon, \tilde{\epsilon})$ as functions of holographic variables.

\paragraph{Perturbative blocks.} Since the accessory parameters depend on the holographic variables only it follows that the $n$-point perturbative block that solves the gradient equations \eqref{BPZ} in the sector of the perturbative operators depends only on the holographic variables, $f^{(1)}= f^{(1)}(w,  w'|\epsilon, \tilde{\epsilon})$, where functions $w,w'$ are defined in \eqref{phv}. On the other hand,  the conformal  transformation \eqref{partial} act on the perturbative conformal block as
\be
\label{con}
f^{(1)}(w,  w'|\epsilon, \tilde{\epsilon}) = f^{(1)}(w|\epsilon, \tilde{\epsilon}) + \sum^{k}_{i=1} \epsilon_i \log w'_i\;,
\ee
where the block on the left-hand side is given in the original $z$-coordinates, while the block on the right-hand side is given in the new $w$-coordinates. Since the accessory parameters are the gradients of the conformal block \eqref{BPZ} then
\be
\ba{c}
\dps
c^{(1)}_i \to \frac{1}{w'_i}\left(c^{(1)}_i- \epsilon_i \frac{w''_i}{w'_i}\right),
\quad
i=1,...,k\;,
\\
\\
\dps
c^{(1)}_j \to c^{(1)}_j\;,
\quad
j = k+1,..., n\;,
\ea
\ee
which is exactly the definition \eqref{def_par}. Indeed, the prefactor in $c_i^{(1)}$ is the Jacobian while the second term in the brackets is the derivative of the $\epsilon_i\log w'_i$.

The gradient equations in the sector of the perturbative operators now read as
\be
\label{efb}
\frac{\partial f^{(1)}(w|\epsilon, \tilde{\epsilon})}{\partial w_i}   = X_i( w|\epsilon, \tilde{\epsilon})\;,
\qquad
i = 1,..., k\;.
\ee
Thus, the conformal block function depends on $n$ independent variables $x = \{z, {\bf z}\}$ only through $k<n$ holographic coordinates $w_i$,   $i=1,...,k$ \eqref{hv}.

\paragraph{Two and more background operators.} In order to move further we recall that up to now all coordinates of the primary operators were kept arbitrary. Now, let the last three coordiniates be fixed, $x_{_{\hspace{-0.5mm}fix}}  = (\hat x_{n-2}$, $\hat x_{n-1}$, $\hat x_n)$. Since we always have two or more heavy background operators, then $x_{_{\hspace{-0.5mm}fix}} = (\hat z_{n-2}, \hat {\bf z}_{n-1}, \hat {\bf z}_{n})$ or
$x_{_{\hspace{-0.5mm}fix}} = (\hat {\bf z}_{n-2}, \hat {\bf z}_{n-1}, \hat {\bf z}_{n})$. In a given OPE channel coordinates $x_m$ with $m\leq n-3$ must be  separated from $x_{_{\hspace{-0.5mm}fix}}$ through a particular OPE ordering.

Supplementing  the monodromy equations \eqref{mon1}, \eqref{mon2} with the linear constraints \eqref{linear1} we obtain the equation system of $k+3$ independent conditions for $n$ accessory parameters. It follows that  the first-order accessory parameters of the background operators remain unfixed by the monodromy equations. In this respect, let us consider two different situations:

\vspace{1mm}

\noindent $\bullet$ Two background operators, i.e. $k=n-2$ perturbative operators. In this case, the equations \eqref{mon2} are absent and we have $n-3$ equations \eqref{mon1} for $n-2$ variables $X_i$, $i=1,...,n-2$. Adding the three constraints \eqref{linear1} along with two accessory parameters of the background operators $Y_{n-1}, Y_{n}$ we obtain in total $n$ equations for $n$ accessory parameters. The three constraints \eqref{linear1} can be solved for three accessory parameters $X_{n-2}$, $Y_{n-1}$, $Y_n$ \eqref{def_par} of the operators located at $x_{_{\hspace{-0.5mm}fix}} = (\hat z_{n-2}, \hat {\bf z}_{n-1}, \hat {\bf z}_{n}) = (z_{n-2},1,\infty)$. We notice that then the three constraints depend on  coordinates $z_i$, $i=1,...,n-2$ only. On the other hand, since the holographic map is invertible (see our comments below \eqref{hv}) we can introduce inverse functions $z_i = z_i(w|{\bf z})$ such that $w_i \circ z_i = 1$. Then, the three constraints can be rewritten in terms of the holographic variables, hence the accessory parameters still depend on the holographic variables only.

\vspace{1mm}

\noindent $\bullet$ For three or more  background operators, i.e. $k\leq n-3$ perturbative operators. In this case, there are exactly $k$ equations \eqref{mon1}, \eqref{mon2} for $k$ variables $X_i$, $i=1,...,k$. The three constraints \eqref{linear1} can be solved for three accessory parameters $Y_m$ \eqref{def_par} with $m = n-2,n-1,n$ of the operators located at $x_{_{\hspace{-0.5mm}fix}} = (\hat {\bf z}_{n-2}, \hat {\bf z}_{n-1}, \hat {\bf z}_{n}) = (0,1,\infty)$. Other parameters $X_i$, $i=1,...,k$ and $Y_j$, $j=k+1,..., n-3$ are independent.
Then, recalling that holographic variables \eqref{phv} are functions of coordinates of the background insertions ${\bf z}_j$ and using \eqref{def_par}, we can evaluate the first derivatives of the perturbative block function  to find the first-order accessory parameters of the background operators,
\be
\label{rem_a_p}
Y_j(w,  w',{\bf z}|\epsilon, \tilde{\epsilon}) = \frac{\partial f^{(1)}(w|\epsilon, \tilde{\epsilon})}{\partial {\bf z}_j} \;,
\qquad
j = k+1,..., n-3\;.
\ee


\paragraph{Uniformization property.} To summarise this section, we  can formulate {\it the uniformization property} of L$^{k}$H$^{n-k}$ perturbative conformal blocks:  in holographic variables, the form of $n$-point block function $f^{(1)}(w|\epsilon, \tilde{\epsilon})$ is defined by $k$ perturbative operators only. In particular, using the holographic parameterization, instead of $n$ equations on the accessory parameters we essentially have $k<n$ equations. So, for instance, the calculation of  the (known) 4-point LLHH block  and $n$-point LLH$^{n-2}$ block is essentially the same and gives the same expression in the holographic variables, see Sections \bref{sec:G43B} and \bref{sec:G5B}. The only difference is that  the holographic function \eqref{phv} is different for different numbers of the background operators so that the perturbative block functions in the $z$-parameterization will be different as well.


\section{Examples with two and three background operators}
\label{sec:E}

In this section, we utilize the holographic variables to work out a few examples of  4-point and 5-point conformal  blocks with two and three background operators. In such cases, the background blocks are the known 2-point and 3-point functions of the operators located in $(1,\infty)$ and $(0,1,\infty)$. Hence, the heavy-light expansion can be elaborated in details and perturbative blocks $f^{(1)}(x|\epsilon, \tilde \epsilon)$ can be  found explicitly.

The zeroth-order solutions to the Fuchsian equation are known for two background operators ($\epsilon_3 = \epsilon_4$ and $({\bf z}_3, {\bf z}_4) = (1,\infty)$) \cite{Fitzpatrick:2014vua}
\be
\label{psi_2}
\psi_{\pm}^{(0)}(y|{\bf z}) = (1-y)^{\frac{1\pm \alpha}{2}}\;,
\ee
where $\alpha = \sqrt{1 - 4 \epsilon_4}$, and for three background operators ($\epsilon_3 \neq  \epsilon_4 = \epsilon_5$  and $({\bf z}_3, {\bf z}_4, {\bf z}_5) =(0,1,\infty)$) \cite{Alkalaev:2019zhs}
\be
\label{psi_3}
\psi_{\pm}^{(0)}(y|{\bf z}) =  (1-y)^{\frac{1+\alpha}{2}} y^{\frac{1\pm\cbeta}{2}}\, {}_2F_1\left(\frac{1\pm\beta}{2},\frac{1\pm\beta}{2}+ \alpha, 1\pm\beta, y\right)\;,
\ee
where $\alpha = \sqrt{1 - 4 \epsilon_4}$ and $\beta = \sqrt{1 - 4 \epsilon_3}$.

The respective holographic functions can be explicitly found
\be
\label{w_2}
w(y|{\bf z}) = (1-y)^{\alpha}\;,
\ee

\be
\label{w_3}
w(y|{\bf z}) =  y^{\beta}\, \frac{\, {}_2F_1\left(\frac{1+\beta}{2},\frac{1+\beta}{2}+ \alpha, 1+\beta, y\right)}{\, {}_2F_1\left(\frac{1-\beta}{2},\frac{1-\beta}{2}+ \alpha, 1-\beta, y\right)}\,.
\ee


\subsection{4-point LLHH conformal block}
\label{sec:G43B}

Here, $x = (z_1, z_2, 1, \infty)$.  In this case, the block is determined by two holographic variables $w_1 = w(z_1)$, $w_2 = w(z_2)$ \eqref{hv}, where $w(y)$ is given by \eqref{w_2}, and two accessory parameters $X_1, X_2$. The monodromy integrals \eqref{GS} read
\be
\ba{c}
\dps \frac{I^{(1)}_{++}}{2 \pi i} =   X_1+ X_2  +\epsilon_1 w_1+ \epsilon_2 w_2 \,, \qquad  \frac{I^{(1)}_{-+}}{2 \pi i} = -(X_1 + X_2) \,,
\\
\\
\dps   \frac{I^{(1)}_{+-}}{2 \pi i} = X_1 w^2_1  + X_2 w^2_2 + 2 \epsilon_1 w_1 +  2 \epsilon_2 w_2\,.
\ea
\ee
The only monodromy equation  \eqref{mon1} in this case reads
\be
\label{m1}
I^{(1)}_{++}I^{(1)}_{++} + I^{(1)}_{+-} I^{(1)}_{-+} = - 4\pi^2 \tilde{\epsilon}_1^2\;.
\ee
Solving the constraints \eqref{linear1} yields the relation, which can be rewritten in the form
\be
\label{m2}
I^{(1)}_{++} = 0\;.
\ee
Thus we have two equations \eqref{m1} and \eqref{m2} for $X_1$ and $X_2$ which are solved as
\be
\label{X4}
\ba{c}
  \dps X_1 = \frac{\epsilon_1 + \epsilon_2}{w_2-w_1} + \frac{\epsilon_2 - \epsilon_1}{2 w_1} + \frac{\sqrt{(\epsilon_1 - \epsilon_2)^2 (w_1-w_2)^2 + 4 \tilde{\epsilon}^2_1 w_1 w_2}}{2 w_1(w_2-w_1)}\;,
  \\
  \\
 \dps X_2 =   \frac{\epsilon_1 + \epsilon_2}{w_1-w_2} + \frac{\epsilon_2 - \epsilon_1}{2 w_2} - \frac{\sqrt{(\epsilon_1 - \epsilon_2)^2 (w_1-w_2)^2 + 4 \tilde{\epsilon}^2_1 w_1 w_2}}{2 w_2(w_1-w_2)}\;,
\ea
\ee
where a sign of the radical term is fixed by the asymptotic behaviour of the resulting conformal block. Integrating the gradient equations \eqref{efb} we find the perturbative 4-point LLHH block function
\be
\label{pb4}
\ba{c}
\dps  f^{(1)}(w| \epsilon, \tilde{\epsilon}) =  - (\epsilon_1 + \epsilon_2) \log (w_1 - w_2)  \\
\\
+ (\epsilon_1 - \epsilon_2)\log \left( (\epsilon_1 - \epsilon_2)(w_1 - w_2) + \sqrt{(\epsilon_1 - \epsilon_2)^2 (w_2 - w_1)^2  + 4 \tilde{\epsilon}^2_1 w_2 w_1 }\right) \\
\\
\dps -\frac{\tilde\epsilon_1}{2}\log\left[\frac{\tilde \epsilon_1 (w_1 + w_2) +  \sqrt {(\epsilon_1 - \epsilon_2)^2 (w_1 -w_2)^2 + 4\tilde{\epsilon}^2_1 w_1 w_2 }}{\tilde \epsilon_1 (w_1 + w_2) -  \sqrt{(\epsilon_1 - \epsilon_2)^2 (w_1 -w_2)^2 + 4\tilde{\epsilon}^2_1 w_1 w_2}}\right]\;.
\ea
\ee
At $\epsilon_1 = \epsilon_2$ it reproduces the block function in the $w$-parametrization  found in \cite{Anous:2019yku}. In particular, the identity block (by definition, $\tilde \epsilon_1 = 0$, whence, $\epsilon_1 = \epsilon_2$) reads
\be
\label{pbv4}
\dps f^{(1)}(w|\epsilon, 0) = - 2 \epsilon_1  \log (w_1 - w_2) \;.
\ee

Going back to the $z$-parameterization by performing the conformal transformation \eqref{con}  we  reproduce the 4-point block functions found in \cite{Fitzpatrick:2014vua,Hijano:2015rla}. E.g., the identity block \eqref{pbv4} will be given by
\be
\label{pbv40}
\dps f^{(1)}(z|\epsilon, 0) = \log\left[\frac{w'(z_1)  w'(z_2)}{(w(z_1) - w(z_2))^2}\right]^{\epsilon_1},
\ee
with the holographic function \eqref{w_2} (the same expression was obtained in \cite{Fitzpatrick:2016mtp} by a different method).

\subsection{4-point LHHH conformal block}
\label{sec:G4B}

Here, $x= (z_1, {\bf z}_2, {\bf z}_3, {\bf z}_4)$. The holographic variable is given by  $w_1 = w_1(z_1)$ \eqref{hv}, where $w(y)$ is given by \eqref{w_3}, and one  accessory parameter $X_1$. The only  monodromy equation  \eqref{mon2}  and its solution read \be
\label{m4}
I^{(1)}_{++} \equiv  X_1 w_1 + \epsilon_1 = 0\;,
\qquad\;
X_1 = - \frac{\epsilon_1}{w_1}\;.
\ee
Also, the background external and intermediate dimensions are restricted by \eqref{fusion}: $\tilde \epsilon_1 = \epsilon_2$.  The respective block is found by integrating \eqref{efb},
\be
\label{4V}
f^{(1)}(w_1|\epsilon, \tilde\epsilon) = - \epsilon_1 \log w_1\;.
\ee
Going back to the $z$-parameterization by using \eqref{w_3} the above function  reproduces the 4-point LHHH block found in \cite{Alkalaev:2019zhs}.

\subsection{5-point  LLHHH conformal block}
\label{sec:G5B}

Here, $x = (z_1, z_2,{\bf z}_3, {\bf z}_4, {\bf z}_5)$. The block is determined by two holographic variables $w_1 = w(z_1)$, $w_2 = w(z_2)$ \eqref{hv}, where $w(y)$ is now given by \eqref{w_3}, and two accessory parameters $X_1, X_2$. The monodromy integrals \eqref{GS} read
\be
\label{S5}
\ba{c}
\dps \frac{I^{(1)}_{++}}{2 \pi i} =   X_1+ X_2  +\epsilon_1 w_1+ \epsilon_2 w_2 \,, \qquad  \frac{I^{(1)}_{-+}}{2 \pi i} = -(X_1 + X_2) \,,
\\
\\
\dps   \frac{I^{(1)}_{+-}}{2 \pi i} = X_1 w^2_1  + X_2 w^2_2 + 2 \epsilon_1 w_1 +  2 \epsilon_2 w_2\,.
\ea
\ee
The monodromy equations \eqref{mon1} and \eqref{mon2} are given by
\be
\label{m5}
I^{(1)}_{++}I^{(1)}_{++} +  I^{(1)}_{+-} I^{(1)}_{-+} = - 4\pi^2 \tilde{\epsilon}_1^2\;,
\qquad
I^{(1)}_{++} =0\;.
\ee
The background external and intermediate dimensions are restricted by \eqref{fusion}: $\tilde \epsilon_2 = \epsilon_3$. The solution to \eqref{m5} is given by
\be
\label{X5}
\ba{c}
  \dps X_1 =  \frac{\epsilon_2 + \epsilon_1}{w_2-w_1} + \frac{\epsilon_2 - \epsilon_1}{2 w_1} + \frac{\sqrt{(\epsilon_1 - \epsilon_2)^2 (w_2-w_1)^2 + 4 \tilde{\epsilon}^2_1 w_1 w_2}}{2 w_1(w_1-w_2)}\;,
  \\
  \\
 \dps X_2 =   \frac{\epsilon_2 + \epsilon_1}{w_1-w_2} + \frac{\epsilon_2 - \epsilon_1}{2 w_2} - \frac{\sqrt{(\epsilon_1 - \epsilon_2)^2 (w_2-w_1)^2 + 4 \tilde{\epsilon}^2_1 w_1 w_2}}{2 w_2(w_2-w_1)}\;.
\ea
\ee
 Integrating the gradient equations \eqref{efb} we find the perturbative 5-point LLHHH block function
\be
\label{pb5}
\ba{c}
\dps  f^{(1)}(w| \epsilon, \tilde{\epsilon}) =  - (\epsilon_1 + \epsilon_2) \log (w_1 - w_2)  \\
\\
+ (\epsilon_1 - \epsilon_2)\log \left( (\epsilon_1 - \epsilon_2)(w_1 - w_2) + \sqrt{(\epsilon_1 - \epsilon_2)^2 (w_2 - w_1)^2  + 4 \tilde{\epsilon}^2_1 w_2 w_1 }\right) \\
\\
\dps -\frac{\tilde\epsilon_1}{2}\log\left[\frac{\tilde \epsilon_1 (w_1 + w_2) +  \sqrt {(\epsilon_1 - \epsilon_2)^2 (w_1 -w_2)^2 + 4\tilde{\epsilon}^2_1 w_1 w_2 }}{\tilde \epsilon_1 (w_1 + w_2) -  \sqrt{(\epsilon_1 - \epsilon_2)^2 (w_1 -w_2)^2 + 4\tilde{\epsilon}^2_1 w_1 w_2}}\right]\;.
\ea
\ee
In particular, the identity block (by definition, $\tilde \epsilon_1 = 0$, whence, $\epsilon_1 = \epsilon_2$) reads
\be
\label{pbv5}
\dps f^{(1)}(w|\epsilon, 0) = - 2 \epsilon_1  \log (w_1 - w_2) \;.
\ee

Note that the above monodromy equations are exactly the same as those in the LLHH case \eqref{m1} and \eqref{m2}, hence, the accessory parameters \eqref{X4} and \eqref{X5} are also the same. In this way, we demonstrate  the uniformization property formulated in the end of Section \bref{sec:HoloV}: in the holographic variables the LLHH block \eqref{pb4} and LLHHH block \eqref{pb5} have the same form. On the other hand,  substituting functions \eqref{w_2} and \eqref{w_3} we will obtain, of course, different block functions in $z$-coordinates.


\section{Conformal blocks as geodesic trees}
\label{sec:bulk}

The holographic function introduced earlier to describe the perturbative blocks also occurs when describing the dual bulk geometry. It allows to identify the dual space as three-dimensional AdS$_3[n-k]$ space with $n-k$ conical singularities created by the background heavy operators. We explicitly show that the 5-point LLHHH perturbative block is calculated by the length of particular geodesic tree in AdS$_3[3]$. The geodesic tree is the same as for the 4-point LLHH perturbative block but in AdS$_3[2]$.


\subsection{Dual geometry}
\label{sec:BM}

Let us consider the three-dimensional metric in the \banados form \cite{Banados:1998gg}
\be
\label{Banados}
ds^2 =  -H(z) dz^2 -\bar{H}(\bar{z})d\bar{z}^2 + \frac{u^2}{4}\,  H(z) \bar{H}(\bar{z}) \, dz d\bar z+ \frac{du^2 + dz d\bar{z}}{u^2}\;,
\ee
where $u\geq 0$, $z, \bar z \in \mathbb{C}$,  and $H, \bar H$ are (anti)holomorphic functions on $\mathbb{C}$
, and the AdS radius is set to one. In the context of the AdS$_3$/CFT$_2$ correspondence, the function $H(z)$ is related to the stress tensor of background operators in $CFT_2$ by
\be
\label{TH}
T(z) = \frac{c}{6} H(z)\;,
\ee
where the central charge $c = 3\rads /2 G_N$ \cite{Brown:1986nw,Balasubramanian:1999re}. Under the boundary conformal transformations $z \rightarrow w(z)$ the stress tensor changes as
\be
\label{trans}
H(z) = \left(w^\prime\right)^2 H(w) + \frac{1}{2}\, \{w,z\}\;,
\qquad
\text{where}
\quad
\{w,z\} =  \frac{w'''}{w'}  - \frac{3}{2}\left(\frac{w''}{w'} \right)^{2}\;.
\ee

The \banados metric \eqref{Banados} can be cast into the Poincare form
\be
\label{PP}
ds^2 = \frac{dv^2+dq d\bar{q}}{v^2}\;,
\ee
with  $v\geq 0$, $q, \bar q \in \mathbb{C}$, by changing the coordinates as follows \cite{Roberts:2012aq}
\be
\label{coot}
\begin{gathered}
q(z,\bar z,u) = w(z)  - \frac{2 u^2 w^\prime(z)^2 \bar w^{\prime\prime}(\bar z)}{4w^\prime(z)\bar w^{\prime}(\bar z)+u^2 w^{\prime\prime}(z)\bar w^{\prime\prime}(\bar z)} \;,
\\
\bar q(z,\bar z,u) = \bar w(\bar z)  - \frac{2 u^2 \bar w^\prime(\bar z)^2 w^{\prime\prime}(z)}{4w^\prime(z)\bar w^{\prime}(\bar z)+u^2 w^{\prime\prime}(z)\bar w^{\prime\prime}(\bar z)} \;,
\\
v(z,\bar z,u) =  u\, \frac{4\left( w^\prime(z)\bar w^\prime(\bar z)\right)^{3/2}}{4w^\prime(z)\bar w^{\prime}(\bar z)+u^2  w^{\prime\prime}(z)\bar w^{\prime\prime}(\bar z)}\;,
\end{gathered}
\ee
where the function $w(z)$ solves the equation  (see \cite{Asplund:2014coa,Fitzpatrick:2015zha,Cresswell:2018mpj,Alkalaev:2019zhs} for more details)
\be
\label{c_sch}
\{w,z\} = \frac{1}{2} H(z)\;.
\ee

It is remarkable that the solution $w(z)$ can be constructed by means of two independent solutions to the auxiliary  Fuchsian equation $\psi''(z)+ H(z)\psi(z) =0$ as
\be
\label{w_dual}
w(z) = \frac{A \psi_1(z) + B \psi_2(z)}{C \psi_1(z) + D \psi_2(z)}\;, \qquad A D  - B C \neq 0\;.
\ee

Identifying the stress tensor of the background operators with the metric-defining function $H(z)$ according to \eqref{TH} we immediately conclude that $\psi_{1,2}(z)$ can be considered as solutions \eqref{sol0} to the auxiliary Fuchsian equation of the monodromy method in the zeroth-order \eqref{fuchs0}. It follows that the mapping function \eqref{w_dual} is exactly the holographic function  \eqref{phv}: its values at points of the boundary primary operators define the holographic variables \eqref{hv}.

Finally, the length of a geodesic stretched between two points $(q_1, \bar{q}_1, v_1)$ and $(q_2, \bar{q}_2, v_2)$ is given by
\be
\label{bg}
\cL = \text{arccosh}\;P = \log (P + \sqrt{P^2 +1})\;, 
\qquad 
P = \frac{(q_1 - q_2)(\bar q_1  - \bar q_2)+ v_1^2 +v_2^2}{2v_1 v_2}\;.
\ee
Here $\cL$ is a real-valued function of endpoint coordinates $(q_i, \bar{q}_i, v_i)$, $i =1, 2$. In the sequel, we will consider geodesic graphs composed  of several geodesic segments lying on the surface $q\bar{q} + v^2 = 1$. In the global coordinates this surface is mapped onto the fixed-time slice.\footnote{Such a condition is convenient, but not necessary. See also Fig. \bref{figure}.} Thus, the total length $\cL$ can be expressed in terms of local coordinates $(\eta,\bar{\eta})$ on this 2-dimensional surface   and factorized into the sum of  holomorphic $\cL(\eta)$ and antiholomorphic $\bar{\cL}(\bar{\eta})$ functions,
\be
\label{hl}
2\cL = \log \left(X(\eta)\bar{X}(\bar{\eta}) \right) \equiv \cL(\eta) + \bar{\cL}(\bar{\eta})\;,
\quad 
\text{where} 
\quad\sqrt{X\bar{X} } \equiv P + \sqrt{P^2 + 1}\;.
\ee 
Finally, note that for a given geodesic graph with a number of boundary attachments the (anti-)holomorphic lengths are functions of boundary endpoint coordinates.


\subsection{Geodesic trees}
\label{sec:GT}

In this section we consider the geometry created by three background operators AdS$_3[3]$ and geodesic trees dual to LHHH and LLHHH perturbative blocks. Since the zeroth-order stress tensor \eqref{TH} has three singular points, then in the \banados coordinates $(z, \bar{z}, u)$ these operators create three singular lines: $(0, 0, u), (1, 1, u)$ and $(\infty, \infty, u)$ stretched along $u\geq 0$.  In the Poincare coordinates, the geometry is completely determined by the properties of the function $w(z)$ given by
\be
\label{dw}
\dps w(z) = z^{\beta}\, \frac{_2F_1\left(\frac{1+\beta}{2},\frac{1+\beta}{2}+ \alpha, 1+\beta, z\right)}{_2F_1\left(\frac{1-\beta}{2},\frac{1-\beta}{2}+ \alpha, 1-\beta, z\right)}\;.
\ee
By construction, this is the same function as \eqref{w_3}. Near the singular points $(0, 1, \infty)$ it can be represented
\be
\label{asymptotics}
\begin{aligned}
z\to 0: \qquad\;& w(z) \sim   z^{\beta}(1+ \cO(z))\;,\\[2pt]
z\to 1: \qquad\;& w(z) \sim  (1-z)^{-\alpha}(1+ \cO(1-z))\;,\\[2pt]
z\to \infty: \qquad & w(z) \sim  z^{-\alpha}(1+ \cO(1/z))\;, \\[2pt]
\end{aligned}
\ee
where $\sim$ means that the coefficients in the Laurent series near these points are omitted. The  function \eqref{dw} is known as the Schwarz triangle function which maps the complex plane $(z, \bar{z})$ onto a curvilinear Schwarz triangle on the plane $(w, \bar{w})$ with vertices at points $w(0), w(1), w(\infty)$ \cite{nehari},
\be
\label{values}
w(0) = 0\;,
\qquad
w(1) = \infty\;,
\qquad
w(\infty) = e^{i \pi \beta} \; \frac{\Gamma(1+\beta) \; \Gamma(\frac{1-\beta}{2} + \alpha) \; \Gamma(\frac{1-\beta}{2})}{\Gamma(1-\beta) \; \Gamma(\frac{1+\beta}{2} + \alpha)\; \Gamma(\frac{1+\beta}{2})}\;.
\ee
The asymptotic behaviour \eqref{asymptotics} suggests that near the singular points corresponding to the background operators the Schwarz triangle describes angle excesses/deficits:
 angle deficits $\beta$ and $\alpha$ at $0$ and $\infty$, an angle excesses $-\alpha$ at $1$.

Let us consider now the singular lines of the background operators in the Poincare coordinates $(q, \bar{q}, v)$. From the asymptotics \eqref{asymptotics} we find that
\be
\ba{l}
z\to 0\;\;:\qquad v(z,\bar z,u) \sim \; u^{-1}(z \bar z)^{\frac{1+\beta}{2}}(1+\cO(z \bar z))\;,

\vspace{2mm}
\\

z\to 1\;\;\;:\qquad v(z,\bar z,u) \sim \;  u^{-1}(\left[(1-z)(1-\bar z)\right]^{\frac{1-\alpha}{2}}(1+\cO((1-z)(1-\bar z))\;,

\vspace{2mm}
\\

z\to \infty\;:\qquad v(z,\bar z,u)  \sim u\; (z\bar z)^{-\frac{1+\alpha}{2}}(1+\cO(1/(z\bar z)))\;.

\ea
\ee
Hence, the singular lines in the \banados coordinates are mapped into the boundary ($v=0$) points $w(0), w(1), w(\infty)$ \eqref{values} in the Poincare coordinates which correspond to the vertices of the Schwartz triangle (see \cite{Alkalaev:2019zhs} for more details).


The general claim of the AdS$_3$/CFT$_2$ correspondence in the large-$c$ regime within the heavy-light expansion reduces to the correspondence formula that relates $n$-point L$^k$H$^{n-k}$ perturbative blocks and holomorphic geodesic lengths 
\be
\label{correspondence}
f_{(k, n-k)}^{(1)}(w|\epsilon, \tilde{\epsilon})\,  =  \,  -\cL_{_{\hspace{-0.5mm}AdS_{_3}[n-k]}}(w|\epsilon, \tilde{\epsilon})\;,
\ee
where a geodesic tree is stretched in the  AdS$_3[n-k]$ space with $n-k$ singularities created by $n-k$ background heavy operators.  The uniformization of perturbative blocks suggests that the form of geodesic trees depends only on the number of perturbative operators. In the bulk, the holographic variables $w_i$, $i=1,...,k$ appear  as coordinates of the boundary attachments of the perturbative operators.

\begin{figure}[H]
  \centering
  \begin{minipage}[h]{0.35\linewidth}
    \includegraphics[width=1\linewidth]{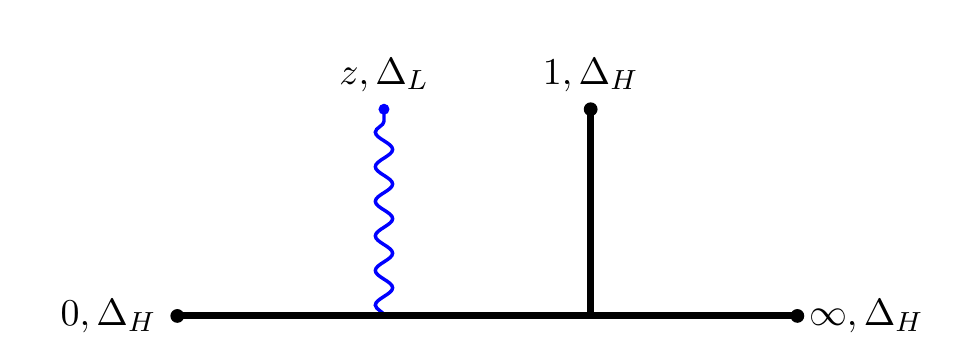}
  \end{minipage}
  \qquad\qquad
   \centering
  \begin{minipage}[h]{0.15\linewidth}
    \includegraphics[width=1\linewidth]{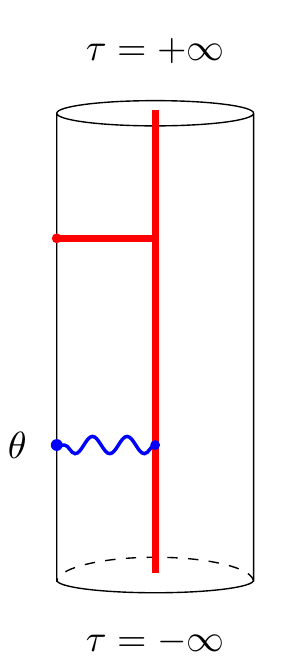}
  \end{minipage}
\caption{The 4-point LHHH block and its holographically dual realization in the three dimensional bulk (a rigid cylinder) global coordinates $\tau \in (-\infty, +\infty)$, $\phi \in [0,2\pi)$, $\rho \in [0, \pi/2)$. The red lines
inside the cylinder visualize  the 3-point function $\langle \cO_H \cO_H \cO_H\rangle$ of heavy operators that created this conical defect
geometry. The wavy blue line denotes the perturbative operator $\cO_L$ propagating in the background. Here  $\theta \equiv  \tau+i \phi =  w_1$. The surface $q\bar q + v^2 = A^2,\; A > 0$ in Poincare coordinates is realized in global coordinates as fixed-$\tau$ slice, $\tau = 2 \log A$. }
\label{figure}
\end{figure}

\paragraph{LHHH block.}  The corresponding  geodesic tree is a line connecting a boundary point $(w_1, \bar w_1, \varepsilon)$ where the cut-off $\varepsilon \to 0$ and the selected bulk point $(0, 0, 1)$ \cite{Alkalaev:2019zhs}. The point is an intersection point of the surface $q \bar{q} + v^2 = 1$  and the line $(0,0,v)$.  This geometrical construction is most manifest in global coordinates where AdS$_3$ is a cylinder. The 2-surface is mapped to a fixed-time slice while the line $(0,0,v)$ is mapped to a line going through the center of cylinder along the time direction (vertical red line $(0,0, \tau)$ on fig. \bref{figure}). Such a line can be visualized as one of legs of the 3-vertex of background operators that created the background geometry.  

Expanding \eqref{bg} in the cut-off parameter we find that according to the (anti-)holomorphic representation \eqref{hl} the length function  is given by  
\be
2\cL \equiv \lengthB(w|\epsilon) + \bar \cL_{_{\hspace{-0.5mm}AdS_{_3}[3]}}(\bar w|\epsilon)  = \epsilon_1 \log w_1 + \epsilon_1 \log \bar w_1 \;.
\ee
Its holomorphic part coincides with the 4-point LHHH perturbative block \eqref{4V}.

\paragraph{LLHHH block.} Let us consider first a geodesic arc stretched between two boundary points $(z_1,\bar z_1, \varepsilon)$ and $(z_2,\bar z_2, \varepsilon)$, where the cut-off $\varepsilon \rightarrow 0$. Expanding the length function \eqref{bg} in the cut-off we find  the weighted length of the arc
\be
\lengthB(w|\epsilon) + \bar \cL_{_{\hspace{-0.5mm}AdS_{_3}[3]}}(\bar w|\epsilon) = 2\epsilon_1 \log (w_2 - w_1)+2\epsilon_1 \log (\bar w_2 - \bar w_1)\;.
\ee
This function (its holomorphic part) coincides with the identity 5-point LLHHH perturbative block given by \eqref{pbv5}.

Now, we consider a geodesic tree with a single trivalent vertex connecting three edges. The two edges are attached to the conformal boundary  at $(w_1, \bar{w}_1, \varepsilon)$ and $(w_2, \bar{w}_2, \varepsilon)$, where the cut-off $\varepsilon \rightarrow 0$, the third edge ends at the selected point $(0,0,1)$ in the bulk. The vertex is the Fermat–Torricelli point $(q, \bar{q}, v)$ which minimizes the corresponding weighted length function.

Then, using \eqref{bg} and condition $q\bar{q} + v^2 = 1$ we compose lengths of three geodesic segments as
\be
\label{length}
\ba{c}
\dps 2 \cL = \sum_{i=1}^2\epsilon_i\log \frac{(q - w_i)(\bar{q} - \bar{w}_i)}{1 - q\bar{q}}
+ \tilde{\epsilon}_1 \log \frac{1 + \sqrt{q\bar{q}}}{1 - \sqrt{q\bar{q}}}\;.
\ea
\ee
Representing $q = t \exp[i \phi]$ and minimizing \eqref{length} with respect to $(t, \phi)$  we find that the Fermat–Torricelli point is given by
\be
t = \frac{\left(4 (w_1 w_2)^{\half} - (\epsilon_1 - \epsilon_2)^2 (w_2 - w_1)^2\right)^{\half} - a_1 (w_2 - w_1)}{ w_2 + w_1 + a_2 (w_2 - w_1)}\;,
\qquad
\cos \phi = \frac{a_3t^2 +2t + a_3}{t^2+2 a_3 t+ 1}\;,
\ee
where
\be
a_1 = (\epsilon_1 + \epsilon_2) a_2 \;,
\qquad
a_2 = \left(\frac{\tilde{\epsilon}^2_1 - (\epsilon_1 - \epsilon_2)^2}{\tilde{\epsilon}^2_1 - (\epsilon_1 +  \epsilon_2)^2}\right)^{\half},
\qquad
a_3 = \frac{\epsilon^2_2-\epsilon^2_1  - \tilde{\epsilon}^2_1}{\epsilon_1 \tilde{\epsilon}_1}\;.
\ee
Substituting these expressions into \eqref{length} we obtain the holomorphic part of the legnth function
\be
\ba{c}
\label{GBWL}
\dps \lengthB(w|\epsilon, \tilde \epsilon) =  (\epsilon_1 + \epsilon_2) \log (w_1 - w_2) \\
\\
- (\epsilon_1 - \epsilon_2)\log \left( (\epsilon_1 - \epsilon_2)(w_1 - w_2) + \sqrt{(\epsilon_1 - \epsilon_2)^2 (w_1 - w_2)^2  + 4 \tilde{\epsilon}^2_1 w_2 w_1 }\right)\\
\\
\dps +\frac{ \tilde\epsilon_1}{2}\log\left[\frac{\tilde \epsilon_1 (w_1 + w_2) +  \sqrt {(\epsilon_1 - \epsilon_2)^2 (w_1 -w_2)^2 + 4\tilde{\epsilon}^2_1 w_1 w_2 }}{\tilde \epsilon_1(w_1 + w_2) -  \sqrt{(\epsilon_1 - \epsilon_2)^2 (w_1 - w_2)^2 + 4\tilde{\epsilon}^2_1 w_1 w_2}}\right]\;,
\ea
\ee
which reproduces the 5-point LLHHH perturbative block \eqref{pb5}.

\section{Summary}
\label{sec:sum}

We showed that using the holographic variables allows to formulate the uniformization property of $n$-point L$^k$H$^{n-k}$ perturbative blocks which claims that their form essentially depends on the number of the perturbative operators $k$ and not on the background operators. In other words, the perturbative conformal block function can be reorganized so that all coordinates are packed into $k$ functions of original coordinates. In this new parameterization, the $n$-point block function has the same form for any given number $k$.

The uniformization property for large-c blocks was originally established for 4-point LLHH blocks \cite{Fitzpatrick:2015zha}, where the coordinate transformation eliminating dependence on the background operators was understood using the standard \cft technique when conformal blocks are represented through matrix elements of Virasoro states.\footnote{In particular, this resulted in the method to calculate large-$c$ perturbative blocks starting from the known $sl(2)$ global blocks \cite{Fitzpatrick:2015zha}   (see also the case of  5-point LLLHH blocks \cite{Alkalaev:2015fbw}).} The uniformizing coordinate transformation in the framework of the monodromy method was considered in \cite{Anous:2019yku} in the case of two background operators. Unlike the discussion in Section \bref{sec:HoloV}, the $w$-coordinate transformation in \cite{Anous:2019yku} is implemented already in the Fuchsian equation so then the monodromy problem is reduced to studying a regularity of the new stress tensor on the $w$-plane. In our case, we follow the standard monodromy analysis noticing that the $z$-dependence can be packed into some new functions and their derivatives \eqref{phv}. A true coordinate change is performed only at the final stage when the the complete set of algebraic equations on the accessory parameters is formulated, see \eqref{con}--\eqref{efb}. Practically, it would be useful to relate two approaches. So, the remarkable form of the monodromy integrals \eqref{GS} suggests that they can be somehow related to regularity of the stress tensor in the new parameterization  (see also our comments below \eqref{GS}).

The holographic variables are indeed holographic as they reappear in the bulk analysis as the boundary coordinates of the perturbative operators in the three-dimensional space AdS$_3[n-k]$ with $n-k$ conical singularities produced by the background operators. From this perspective, the uniformization property is more obvious  because it is quite natural that $k$ perturbative operators produce the same geodesic tree  no matter how many background operators created the bulk space. In fact, the background operators with dimensions $\Delta < c/24$ (cf. \eqref{alpha}) produce conical singularities so that AdS$_3[n-k]$ is locally AdS$_3$. By casting the original \banados metric to the Poincare form, all dependence on  positions of the background operators is now hidden inside the mapping function and its domain of definition. It turns out that the same function defines the holographic coordinates in the boundary CFT$_2$ because of the same Fuchsian equation that underlies both bulk and boundary calculations. Here, the Schwarz triangle function \eqref{dw} which is the mapping function in  AdS$_3[3]$ and  the holographic function \eqref{w_3} in \cft clearly illustrates all details.  

We have explicitly demonstrated this machinery for LLHH and LLHHH perturbative blocks. Going beyond more than three background operators faces the problem of lacking explicit expressions for higher-point conformal blocks\footnote{For recent study of higher-point conformal blocks, see   \cite{Rosenhaus:2018zqn,Parikh:2019ygo,Jepsen:2019svc,Parikh:2019dvm}.} that define the background part H$^{n-k}$ of the original $n$-point large-$c$  block function. Nonetheless, the uniformization property claims that the perturbative L$^k$ part will be the same.

The large-$c$ multi-point Virasoro blocks considered in the present paper can be used in the study of many interesting physics problems. Here, the main up-to-date application can be found in analyzing  the entanglement entropy that according to \cite{Calabrese:2004eu,Calabrese:2009qy} basically reduces to calculating higher-point correlators of heavy operators ($\Delta = \cO(c)$). For  recent studies of large-$c$ conformal blocks in this context see, e.g.  \cite{Hartman:2013mia,Banerjee:2016qca, Anous:2019yku}. Another possible application is related to the quantum chaos and its characterization known as out-of-time ordered correlators (OTOC) (see, e.g. recent \cite{Kusuki:2019gjs} and references therein).  Also, let us note that the monodromy approach in terms of holographic variables can be applied  to systems enjoying large-$N$ expansion and symmetries other than Virasoro. For example, it would be interesting to identify holographic variables for BMS$_3$ blocks considered in large-$c$ regime in the context of the flat-space holography (see e.g. \cite{Bagchi:2016geg,Hijano:2019qmi,Hijano:2018nhq}).

\vspace{4mm}

\noindent \textbf{Acknowledgements.} The work was supported by the RFBR grant No 18-02-01024  and by the Foundation for the Advancement of Theoretical Physics and Mathematics “BASIS”.

\providecommand{\href}[2]{#2}\begingroup\raggedright\endgroup
\end{document}